\documentclass{article}

\usepackage{spconfa4,amsmath,graphicx}

\usepackage{amsfonts,amssymb,bm,empheq,nicefrac,nicematrix,optidef,upgreek}

\usepackage{stfloats}
\usepackage[subrefformat=parens,labelformat=parens,caption=false]{subfig}

\usepackage[nopatch=footnote]{microtype}
\usepackage{dirtytalk,siunitx,soul,tcolorbox,textcomp,lipsum}

\usepackage[subtle,title=normal,sections=normal,margins=normal,bibliography=normal,mathdisplays=normal,mathspacing=normal]{savetrees}

\usepackage{cite}
\usepackage[dvipsnames]{xcolor}
\usepackage{xurl}

\usepackage[
  colorlinks=true,
  linkcolor=blue,
  citecolor=blue,
  urlcolor=blue,
  filecolor=blue,
  unicode=true,
  bookmarksnumbered=true,
  bookmarksopen=true,
  pdfpagemode=UseOutlines,
  pdftitle={Robust Soft-Constrained Spatially Selective Active Noise Control for Hearables Under Secondary Path Variations},
  pdfauthor={Tong Xiao, Reinhild Roden, Matthias Blau, Simon Doclo},
  pdfkeywords={active noise control, spatially selective active noise control, soft constraints, secondary-path variations, hearables}
]{hyperref}

\usepackage[capitalize,nameinlink]{cleveref}
\crefname{equation}{}{}
\Crefname{equation}{Equation}{Equations}

\usepackage{etoolbox}

\expandafter\def\expandafter\normalequation\expandafter{\equation}
\expandafter\def\expandafter\endnormalequation\expandafter{\endequation}

\renewenvironment{equation}{\abovedisplayskip=2pt\belowdisplayskip=2pt\normalequation}{\endnormalequation}

\AtBeginEnvironment{align}{%
  \setlength{\abovedisplayskip}{5pt plus 1pt minus 2pt}%
  \setlength{\belowdisplayskip}{5pt plus 1pt minus 2pt}%
  \setlength{\abovedisplayshortskip}{5pt plus 1pt minus 2pt}%
  \setlength{\belowdisplayshortskip}{5pt plus 1pt minus 2pt}%
  \setlength{\jot}{3pt}%
}

\title{ROBUST SOFT-CONSTRAINED SPATIALLY SELECTIVE ACTIVE NOISE CONTROL FOR HEARABLES UNDER SECONDARY PATH VARIATIONS}

\name{%
  Tong Xiao$^{1}$,%
  \thanks{This research was funded by the Deutsche Forschungsgemeinschaft (DFG, German Research Foundation) -- Project-ID 352015383 -- SFB 1330 C1, and Germany's Excellence Strategy -- EXC 2177/2 -- Project ID 390895286.}
  Reinhild Roden$^{2}$,
  Matthias Blau$^{2}$,
  Simon Doclo$^{1}$%
}

\address{%
  $^1$ Department of Medical Physics and Acoustics and Cluster of Excellence \say{Hearing4all.connects},\\
  Carl von Ossietzky Universit\"{a}t Oldenburg, Germany\\
  tong.xiao@uni-oldenburg.de, simon.doclo@uni-oldenburg.de\\
  $^2$ Institut f\"{u}r H\"{o}rtechnik und Audiologie and Cluster of Excellence \say{Hearing4all.connects},\\
  Jade Hochschule, Oldenburg, Germany\\
  reinhild.roden@jade-hs.de, matthias.blau@jade-hs.de\\
}

\begin{document}

\ninept
\maketitle

\begin{abstract}
Spatially selective active noise control (SSANC) hearables aim to attenuate noise from certain directions at the eardrum while preserving desired speech arriving from selected directions. Existing SSANC systems typically assume an accurate estimate of the secondary path from the loudspeaker to the inner error microphone. In practice, however, this path varies across users and device fits, which can degrade performance and compromise system stability. This paper proposes a robust soft-constrained optimization framework that computes a single control filter by minimizing the average cost over a set of secondary-path estimates derived from human measurements. Simulations show that the proposed approach achieves slightly lower mean performance than the matched case but substantially narrows the performance spread under secondary-path mismatch. Real-time experiments using the tested head-and-torso simulator show good agreement with the simulations.
\end{abstract}


\section{Introduction}
\label{sec:intro}

Active noise control (ANC) hearables use secondary sources to generate anti-noise to minimize the noise leakage at the eardrum \cite{Kuo1996active, Elliott2000, Hansen2012active}. Conventional ANC hearables treat all incoming sounds as noise \cite{benois2022optimization, Hilgemann2024data}. This becomes problematic in acoustic environments when desired speech is present \cite{Chang2016Listening, gupta2022augmented, Serizel2010integrated, Dalga2013influence, Patel2020design, xiao2023spatial}. In these cases, the user may want to focus on a specific speech source from a certain direction (e.g., the front) while reducing the noise leakage from other directions.

To achieve this, modern hearables employ a combination of outer microphones and inner error microphones. While traditional beamforming methods preserve desired speech, they often ignore the leakage at the inner error microphones \cite{vanveen1988, gannot2017consolidated, Doclo2015multichannel}. Spatially selective ANC (SSANC) combines beamforming and ANC principles to preserve speech from desired directions while reducing noise from other directions \cite{xiao2023spatial, Xiao2024icassp, Xiao2025fa}. A soft-constrained SSANC system was recently proposed to balance noise reduction and speech distortion using a trade-off parameter \cite{Xiao2025soft}. So far, these approaches have assumed an accurate secondary path estimate between the secondary source and the inner error microphone. In practice, these paths vary across individuals due to ear anatomy and device fit. Such mismatches can degrade performance and compromise system stability.

In this paper, we propose a robust optimization framework for soft-constrained SSANC under secondary-path variations. We assume that a set of secondary-path estimates is available to represent variations across different users and device fittings when the exact secondary path is unknown. Instead of estimating a specific individual path, we determine a control filter that is robust to secondary-path variations by minimizing the average cost function evaluated over this set of estimates. A related averaging method has previously been applied to sound pressure equalization~\cite{schepker2022robust}. Here, we derive and experimentally validate its application to soft-constrained SSANC, where secondary-path variations influence both the noise-reduction and desired-speech-preservation objectives. For experimental validation, we implemented the proposed framework in real time using a dSPACE SCALEXIO LabBox system. The experimental results for the tested configuration are consistent with the simulation findings. This framework provides a practical design option for hearables when accurate secondary-path estimates are unavailable.

\section{Signal model}
\label{sec:time_domain_signal_model}

\begin{figure}[t]
    \centering
    \includegraphics[width=0.78\linewidth]{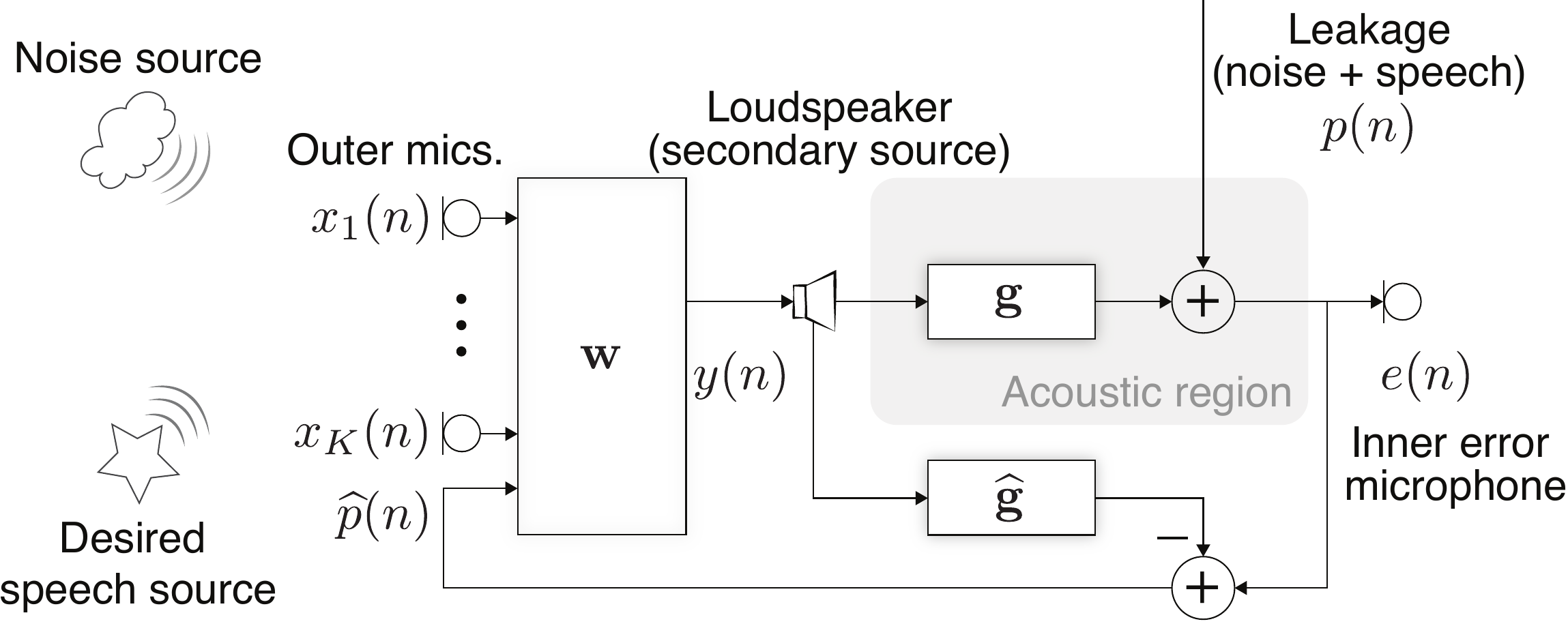}
    \vspace{-10pt}
    \caption{Block diagram of an SSANC system with $K$ outer microphones, one inner error microphone and one loudspeaker (i.e., secondary source). The control filter is denoted by $\mathbf{w}$, the secondary path is denoted by $\mathbf{g}$, and its estimate by $\widehat{\mathbf{g}}$.}
    \label{fig:ssanc}
    \vspace{-10pt}
\end{figure}

As shown in~\cref{fig:ssanc}, we consider a hearable with $K$ outer microphones. Without loss of generality, we consider one loudspeaker as the secondary source and one inner error microphone, resulting in a total of $K+1$ microphones. 
We assume that the acoustic feedback paths between the loudspeaker and the outer microphones are known, such that acoustic feedback can be canceled. 

The inner error microphone signal $e(n)$, where $n$ denotes the discrete-time index, is given by
\begin{equation}
        e(n) = p(n) + ( \mathbf{G}\mathbf{w})^\mathcal{T}{\mathbf{x}}(n) , \label{eq:e_dGwx}
\end{equation}
where $(\cdot )^\mathcal{T}$ denotes the transpose operator. The leakage (including noise and desired speech) at the inner error microphone is denoted by $p(n)$, and the anti-noise component at the inner error microphone is given by $( \mathbf{G}\mathbf{w})^\mathcal{T}{\mathbf{x}}(n)$, where ${\mathbf{w}}$ is the stacked control filter, ${\mathbf{x}}(n)$ is the stacked input vector, and $\mathbf{G}$ represents the secondary path convolution matrix.
The stacked control filter ${\mathbf{w}}$ is defined as
\begin{subequations}
\begin{align}
    \mathbf{w} &= [ \mathbf{w}^\mathcal{T}_1 \;\;\; \mathbf{w}^\mathcal{T}_2 \;\;\; \dots \;\;\; \mathbf{w}^\mathcal{T}_{K+1} \, ]^\mathcal{T} \in \mathbb{R}^{(K+1)L_w} ,
    \\
    \mathbf{w}_k &= \left[{w_{k,0}} \,\,\, {w_{k,1}} \,\, \dots \ {w_{k,{L_w}-1} } \right]^\mathcal{T} \in \mathbb{R}^{L_w},  
\end{align}
\end{subequations}
where $L_w$ denotes the control filter length for each channel. 
The convolution matrix of the secondary path $\mathbf{g} = [ {g}_0\ {g}_1\ \allowbreak \dots\ \allowbreak {g}_{L_g-1} ]^\mathcal{T}$ with a length of $L_g$ is defined as
\begin{subequations}
\begin{align}
    \mathbf{G} &= \mathrm{blkdiag}\left({\mathsf{G}}  \dots  {\mathsf{G}} \right) \in \mathbb{R}^{(K+1)L \times (K+1)L_w},  \label{eq:G_tilde_multi}
\\ 
    {\mathsf{G}} &= 
        \begin{bNiceMatrix}
                g_0       &  \cdots  & 0         \\[-4pt]
                \vdots    &  \ddots  & \vdots    \\[-4pt]
                g_{L_g-1} &  \ddots  & g_0       \\[-4pt]
                \vdots    &  \ddots  & \vdots    \\[-2pt]
                0         &  \cdots  & g_{L_g-1} \\[0pt]
        \end{bNiceMatrix}
        \in \mathbb{R}^{L \times L_w} ,
    \label{eq:G_hat}
\end{align}
\end{subequations}
where $L=L_g+L_w-1$.
As input signals to the control filter, we consider the $K$ outer microphone signals $\mathbf{x}_k(n)$, $k=1, \dots , K$, and an estimate of the leakage $\widehat{\mathbf{p}}(n)$. The stacked input vector ${\mathbf{x}}(n)$ is defined as
\begin{equation}
    {\mathbf{x}}(n) = [\mathbf{x}_{1}^\mathcal{T}(n) \, \dots \, \mathbf{x}_{K}^\mathcal{T}(n) \,\, \widehat{\mathbf{p}}^\mathcal{T}(n) ]^\mathcal{T} \in \mathbb{R}^{(K+1)L} , \label{eq:x_tilde_multi}
\end{equation}
with
\begin{subequations}
\begin{align}
    \mathbf{x}_k(n) &= \left[ x_k(n) \ \dots \ x_k(n-L+1) \right]^\mathcal{T} \in \mathbb{R}^{L} ,  \label{eq:x_k_vec}
\\ 
    \widehat{\mathbf{p}}(n) &= \left[ \,\, \widehat{p}\,(n) \, \ \dots \ \,\, \widehat{p}\,(n-L+1) \right]^\mathcal{T} \in \mathbb{R}^{L}. \label{eq:p_vec}
\end{align}
\end{subequations}
The estimated leakage $\widehat{p}(n)$ can be computed from the inner error microphone signal $e(n)$ and the loudspeaker signal vector $\mathbf{y}(n)=[y(n)\ \dots\ y(n-L_g+1)]^\mathcal{T}$, where $y(n)$ is obtained by filtering each of the $K+1$ input signals with its corresponding control filter $\mathbf{w}_k$ and summing the resulting outputs, and an estimate of the secondary path $\widehat{\mathbf{g}}=[\widehat{g}_0\ \widehat{g}_1\ \allowbreak \dots\ \allowbreak \widehat{g}_{L_g-1}]^\mathcal{T}\in\mathbb{R}^{L_g}$, as
\begin{equation}
\widehat{p}(n)=e(n)-\widehat{\mathbf{g}}^\mathcal{T} \mathbf{y}(n) .
\label{eq:d_hat}
\end{equation}
To derive the optimal control filter, we first assume an accurate secondary path estimate, $\widehat{\mathbf{g}} = \mathbf{g}$. Under this assumption, the estimated leakage equals the true leakage, $p(n) = \widehat{p}(n) = \mathbf{q}^\mathcal{T}\mathbf{x}(n)$, with 
\begin{subequations}
\begin{align}
    \mathbf{q} &= [ \, \mathbf{0}^\mathcal{T} \ \ldots \ \mathbf{0}^\mathcal{T} \;\; \bm{\updelta}^\mathcal{T} ]^\mathcal{T} \in \mathbb{R}^{(K+1)L} ,    \label{eq:delta_tilde_multi}
\\
    \bm{\updelta} &= \left[ \, 1 \;\;\;\; 0 \;\;\; \dots \;\;\; 0 \;\; \right]^\mathcal{T}  \in \mathbb{R}^{L} .
\end{align}
\end{subequations}
Hence, the inner error microphone signal in~\labelcref{eq:e_dGwx} can be rewritten as
\begin{equation}
    \boxed{
        e(n) = (\mathbf{q} + \mathbf{G}\mathbf{w})^\mathcal{T}{\mathbf{x}}(n) .
        }
        \label{eq:e_qGwx} 
\end{equation}

While this assumption is used to derive the optimal control filter, the true secondary path can vary for different individuals and for different device fits. This mismatch can cause the estimated leakage to differ from the true leakage, which impacts the system performance. However, we first establish a matched-case performance reference before analyzing secondary-path mismatch.

\section{SSANC formulation and evaluation cases}
\label{sec:ssanc_formulation}

The objective of the SSANC system is to minimize the power of the inner error microphone signal while preserving the delayed desired speech component of a chosen reference microphone signal. A soft-constrained optimization has been proposed to balance the noise reduction and the speech distortion \cite{Xiao2025soft}. The cost function is formulated as
\begin{align}
    \min_{\mathbf{w}} \, \mathcal{E}\{e^2(n)\} + \mathbf{w}^\mathcal{T} \mathbf{B} \mathbf{w} + \mu \| \mathbf{H}(\mathbf{q} + \mathbf{G}\mathbf{w}) - \alpha \bm{\updelta}_{\Delta} \|_2^2 ,
    \label{eq:cost_function}
\end{align}
where $\mathcal{E}\{\cdot\}$ denotes the mathematical expectation operator. $\mathbf{w}^\mathcal{T} \mathbf{B} \mathbf{w}$ is a regularization term to prevent overloading the secondary source and to reduce the risk of instability, with $\mathbf{B}$ being a block-diagonal matrix applying different penalty weights to the feedforward (FF) and feedback (FB) channels. It is defined as 
\begin{equation}
    \hspace{-2pt} \mathbf{B} \!=\! \mathrm{blkdiag}\left( \beta_{\mathrm{FF}}\mathbf{I}, \dots, \beta_{\mathrm{FF}}\mathbf{I}, \beta_{\mathrm{FB}}\mathbf{I} \right) \! \in \! \mathbb{R}^{(K+1)L_w \! \times \! (K+1)L_w}\!,
\end{equation}
where $\beta_{\mathrm{FF}}$ and $\beta_{\mathrm{FB}}$ are the regularization parameters for the FF and FB channels, respectively, and $\mathbf{I}$ denotes the $L_w \times L_w$ identity matrix.
The real-valued positive parameter $\mu$ controls the trade-off between noise reduction and speech distortion. A larger $\mu$ emphasizes speech preservation, while a smaller $\mu$ allows for more noise reduction at the cost of increased speech distortion. 
The matrix $\mathbf{H}$ contains the acausal relative impulse responses (ReIRs), and $\bm{\updelta}_{\Delta}$ represents the delayed target response with a delay of $\Delta$ samples \cite{Xiao2024icassp, Xiao2025fa},
\begin{subequations}
    \begin{align}
        \hspace*{-0.6em} \mathbf{H} &= \left[\mathbf{H}_1 \ \dots \ \mathbf{H}_{K+1} \right] \in \mathbb{R}^{(L_a+L_h+L-1) \times (K+1)L}  \label{eq:H_time}, \\
        \hspace*{-0.6em} \mathbf{H}_k &=  \!
            \begin{bNiceMatrix}
            h_{k,-L_a} & \!\! \cdots   \!\!  & 0                 \\[-4pt]
            \vdots     & \!\! \ddots   \!\!  & \vdots            \\[-4pt]
            h_{k,L_h-1}& \!\! \ddots   \!\!  & h_{k,-L_a}        \\[-4pt]
            \vdots     & \!\! \ddots   \!\!  & \vdots            \\[-2pt]
            0          & \!\! \cdots   \!\!  & h_{k,L_h-1}       \\[0pt]
            \end{bNiceMatrix}
            \in \mathbb{R}^{(L_a+L_h+L-1) \times L}  , \label{eq:Hk_time} \\
        \hspace*{-0.6em} \bm{\updelta}_{\Delta} &= [ \, \underbrace{0 \, \dots \, 0}_{L_a} \, \underbrace{0 \, \dots \, 0}_{\Delta}  \underbrace{1 \; 0 \, \dots \, 0  }_{L_h+L-1-\Delta} ]^\mathcal{T} \in \mathbb{R}^{L_a+L_h+L-1}  , \label{eq:delta_Delta}
    \end{align}
\end{subequations}
where $\mathbf{H}_{k}$ is the convolution matrix of the ReIR for the $k$-th channel with respect to the chosen reference microphone, with $L_a$ and $L_h$ denoting the length of the anti-causal and causal parts of the ReIR, respectively~\cite{Xiao2025fa}. $\alpha$ is a real-valued positive amplification factor of the desired speech signal. It should be particularly noted that although the ReIRs are acausal, the control filter $\mathbf{w}$ for reducing noise is still causal.

The solution to \labelcref{eq:cost_function} is found to be~\cite{Xiao2025soft}
\begin{equation}
    \hspace*{-6pt}
    \mathbf{w}_\mathrm{soft} \! =  
        \!-\! \left( \bm{\Phi}_{\mathbf{rr}} \! + \! \mu \mathbf{G}^\mathcal{T} \! \mathbf{H}^\mathcal{T} \! \mathbf{H} \mathbf{G} \right)^{-1} \!\!
        \left[ \! \bm{\upphi} \!-\! \mu \mathbf{G}^\mathcal{T} \! \mathbf{H}^\mathcal{T} \! ( \alpha \bm{\updelta}_{\Delta} \!\!-\! \mathbf{H} \mathbf{q}) \right] \!\!  , \!\!
    \label{eq:w_ssanc_time_soft}
\end{equation}
where 
\begin{subequations}
\label{eq:system_matrices}
\vspace{-4pt}
\begin{align}
    \bm{\Phi}_{\mathbf{rr}} &= \mathbf{G}^\mathcal{T} \mathcal{E} \{\mathbf{x}(n)\mathbf{x}^\mathcal{T}(n)\} \mathbf{G} + \mathbf{B}, \label{eq:Phirr_B} \\
    \bm{\upphi} &= \mathbf{G}^\mathcal{T} \mathcal{E} \{\mathbf{x}(n)\mathbf{x}^\mathcal{T}(n)\} \mathbf{q}. \label{eq:phi}
\end{align}
\end{subequations}

\subsection{Case 1: Matched case (oracle)}
\label{ssec:case1}
To establish a performance bound, we first consider the matched scenario where the secondary path estimate used for optimization exactly matches the physical secondary path. In this case, the control filter in \labelcref{eq:w_ssanc_time_soft} is both optimized and evaluated using the same secondary path. This scenario serves as an oracle performance level and represents an approximate upper bound within the considered formulation.

\subsection{Case 2: Mismatched case}
\label{ssec:case2}
In practice, secondary-path variations may arise. To analyze the sensitivity of the control filter optimization to secondary-path mismatch, we consider a set of $J$ secondary-path estimates $\{ \mathbf{G}_j \}_{j=1}^{J}$, which represent a range of conditions that may or may not accurately characterize the true secondary path $\mathbf{G}$.

For each estimate, a control filter $\mathbf{w}_j$ is calculated based on the estimate $\mathbf{G}_j$ as
\begin{equation}
    \mathbf{w}_j \!=\! - (\bm{\Phi}_{\mathbf{rr},j} + \mu \mathbf{G}_j^\mathcal{T} \mathbf{H}^\mathcal{T} \mathbf{H} \mathbf{G}_j)^{-1} [\bm{\upphi}_j - \mu \mathbf{G}_j^\mathcal{T} \mathbf{H}^\mathcal{T} (\alpha \bm{\updelta}_{\Delta} - \mathbf{H} \mathbf{q})] ,
\end{equation}
where the mismatched correlation matrix $\bm{\Phi}_{\mathbf{rr},j}$ and vector $\bm{\upphi}_j$ are constructed according to \labelcref{eq:system_matrices} by replacing $\mathbf{G}$ with $\mathbf{G}_j$. The input correlation matrix remains fixed. Each $j$-th secondary path estimate is used for optimization, and then the remaining $J-1$ paths are used for evaluation. This case illustrates the distribution of performance under various degrees of estimate accuracy.

\section{Case 3: Robust optimization}
\label{sec:robust_optimization}

Significant secondary-path variations can lead to performance degradation when the control filter is optimized only for a single nominal secondary path. To ensure consistency across a diverse set of conditions, we propose a robust optimization framework that minimizes the average cost over a set of secondary-path estimates.

The objective is to determine a robust control filter $\mathbf{w}_{\mathrm{robust}}$ that minimizes the average cost over the set of $J$ secondary-path estimates, which is formulated as
\begin{equation}
    \boxed{
    \min_{\mathbf{w}} \, \frac{1}{J} \sum_{j=1}^{J} \left[ \mathcal{E}\{e_j^2(n)\} + \mu \| \mathbf{H}(\mathbf{q} + \mathbf{G}_j \mathbf{w}) \! - \! \alpha \bm{\updelta}_{\Delta} \|_2^2 \right] \! + \! \mathbf{w}^\mathcal{T} \mathbf{B} \mathbf{w} ,
    }
\end{equation}
where $e_j(n)$ denotes the inner error microphone signal associated with the $j$-th secondary path estimate. The robust filter $\mathbf{w}_{\mathrm{robust}}$ can be derived as
\begin{empheq}[box=\fbox]{multline}
    \vspace{0pt}
    \mathbf{w}_{\mathrm{robust}} = - \Big( \overline{\bm{\Phi}}_{\mathbf{rr}} + \mu \frac{1}{J} \sum_{j=1}^{J} \mathbf{G}_j^\mathcal{T} \mathbf{H}^\mathcal{T} \mathbf{H} \mathbf{G}_j \Big)^{-1} \\
    \times \Big[ \overline{\bm{\upphi}} - \mu \frac{1}{J} \sum_{j=1}^{J} \mathbf{G}_j^\mathcal{T} \mathbf{H}^\mathcal{T} (\alpha \bm{\updelta}_{\Delta} - \mathbf{H} \mathbf{q}) \Big] ,
    \label{eq:w_robust_solution}
\end{empheq}
where 
\begin{subequations}
\label{eq:avg_system_matrices}
\begin{align}
    \overline{\bm{\Phi}}_{\mathbf{rr}} &= \frac{1}{J} \sum_{j=1}^{J} \mathbf{G}_j^\mathcal{T} \mathcal{E} \{\mathbf{x}(n)\mathbf{x}^\mathcal{T}(n)\} \mathbf{G}_j + \mathbf{B}, \label{eq:Phirr_avg} \\
    \overline{\bm{\upphi}} &= \frac{1}{J} \sum_{j=1}^{J} \mathbf{G}_j^\mathcal{T} \mathcal{E} \{\mathbf{x}(n)\mathbf{x}^\mathcal{T}(n)\} \mathbf{q}. \label{eq:phi_avg}
\end{align}
\end{subequations}
By averaging the path-dependent cost functions over the set of secondary-path estimates, the proposed formulation explicitly accounts for the individual path variations and is expected to reduce sensitivity to mismatch. This differs from first averaging the secondary paths and optimizing for the resulting nominal path, which may not correspond to any physically occurring device fit and may mask magnitude and phase variations that are relevant to robustness and stability.

\section{Evaluation}
\label{sec:evaluation}

\subsection{Setup}
\begin{figure}[t]
    \centering
    \captionsetup[subfloat]{captionskip=0pt,farskip=0pt}
    \subfloat[]{\includegraphics[height=0.32\linewidth]{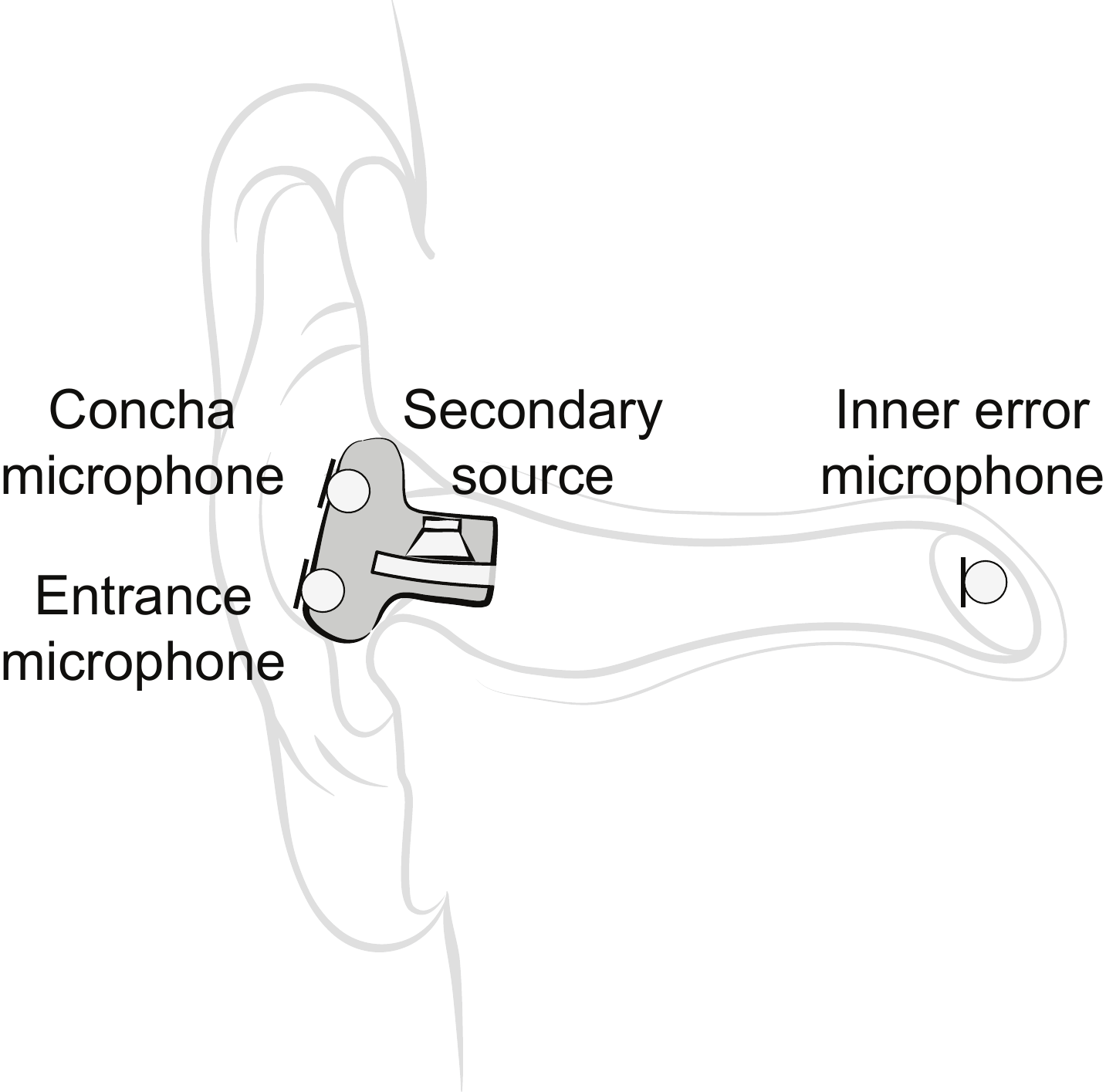}  \label{fig:ear}}
    \qquad  
    \subfloat[]{\includegraphics[height=0.32\linewidth]{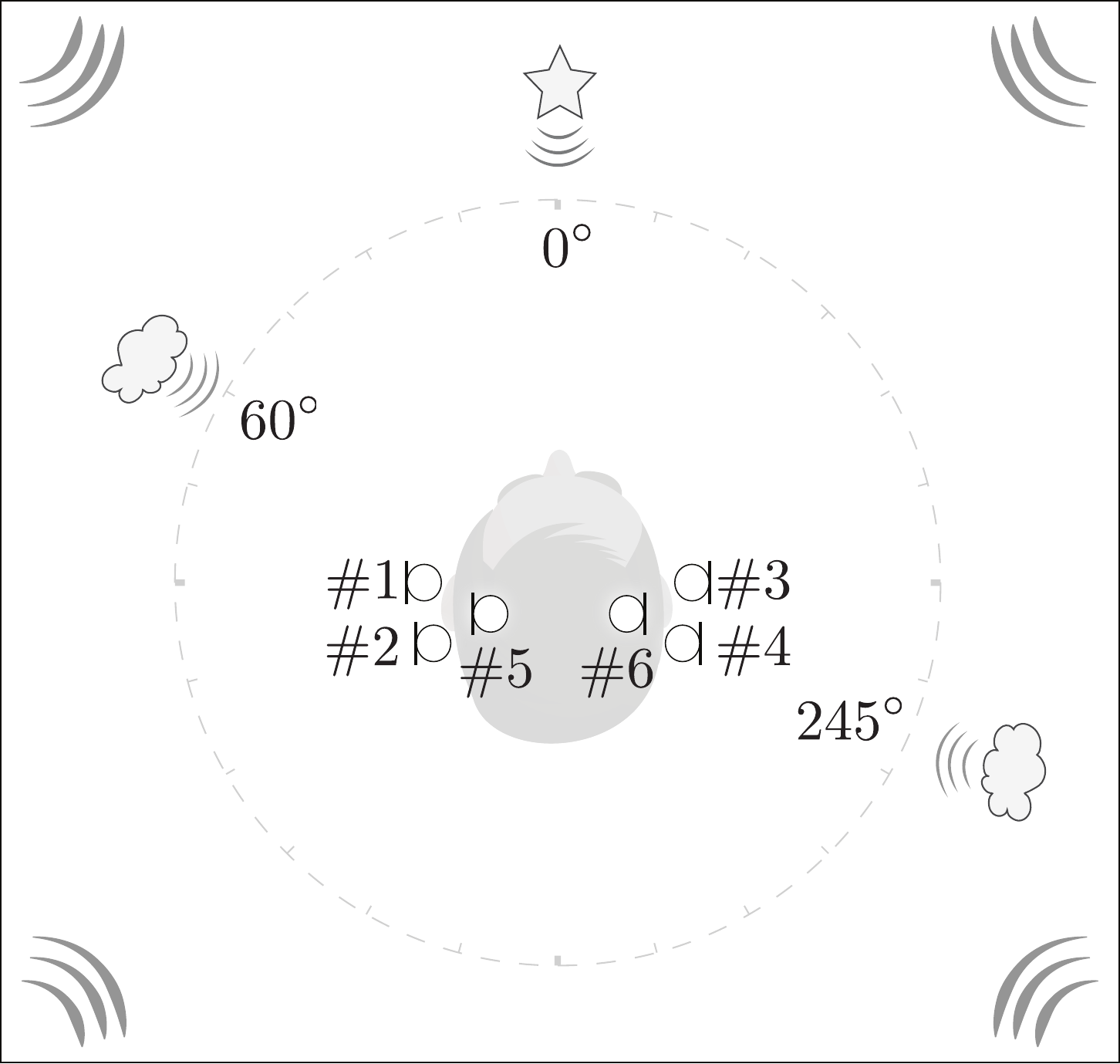}  \label{fig:scene}}
    \vspace{-8pt}
    \caption{(a) Illustration of the closed-fitting hearable. (b) Acoustic scenario with one desired speech source at $0^\circ$, two noise sources at $60^\circ$ and $245^\circ$, and environmental noise from 12 loudspeakers (omitted for clarity).}
    \vspace{-8pt}
    \label{fig:setup}
\end{figure}

For the evaluation, we considered a pair of closed-fitting hearables inserted into both ears of a GRAS 45BB-12 KEMAR Head \& Torso simulator, as illustrated in \cref{fig:ear}. We used four outer microphones (entrance microphones and concha microphones at the left and right ears, labeled as \#1--\#4), two inner error microphones (located at the eardrum, labeled as \#5 and \#6), and the inner drivers acting as the secondary sources. Control was performed independently for each side, using the four outer microphones (\#1--\#4) in combination with the respective inner error microphone (\#5 or \#6). The entrance microphones \#1 and \#3 were chosen as the reference microphones due to the clear speech pickup.

The acoustic scenario (\cref{fig:scene}) was set up in a moderately reverberant room ($7 \times 6 \times 2.7~\mathrm{m}$, $T_{60} \approx 370~\mathrm{ms}$). The desired speech source ($0^\circ$, \qty{0.7}{\meter}) consisted of VCTK~\cite{veaux2017cstr} utterances ``005--006'' from speaker ``p361''. Two airplane cabin noise sources~\cite{BBCSoundEffects07025055} were positioned at $60^\circ$ (\qty{0.7}{\meter}) and $245^\circ$ (\qty{0.9}{\meter}). Additionally, a 12-loudspeaker circular array (radius of \qty{1.8}{\meter}) generated a diffuse pub scene rendered by TASCAR~\cite{grimm2019toolbox, grimm2021pub}. All signals were recorded for \qty{10}{\second} at \qty{40}{\kilo\hertz} using a dSPACE SCALEXIO LabBox with a field-programmable gate array (FPGA). The average leakage signal-to-noise ratio (SNR) at the inner error microphones was \qty{-7.0}{\decibel}, with \qty{-6.6}{\decibel} at the left ear and \qty{-7.2}{\decibel} at the right ear.

The secondary paths for KEMAR were identified by least-squares estimation under white noise excitation. To model secondary-path variations, we derived complex human-to-KEMAR relative frequency responses from measurements~\cite{denk2019one,denk2021hearpiece}. The corresponding variation filters were obtained by inverse Fourier transformation and subsequently truncated to a short causal interval and tapered with a Hann window before being applied to the measured KEMAR secondary path, yielding a set of $J=44$ secondary paths. In the simulations, this set of $J$ paths was used to define the three evaluation cases. In Case~1 (Matched), each path was optimized and evaluated on itself, resulting in $J$ scenarios. In Case~2 (Mismatched), each $j$-th path was used for optimization and evaluated on the remaining $J-1$ paths, yielding $J(J-1)$ scenarios in total. In Case~3 (Robust), a single robust filter was optimized using all $J$ paths and evaluated across the same set.

The ReIRs were assumed to be known and derived by first obtaining absolute impulse responses from the desired source to all microphones via exponential sine sweep (ESS). A least-squares estimation, using white noise as a synthetic excitation, was then applied to identify the ReIRs with respect to the reference microphone.

The lengths of the control filters and secondary paths were set to $L_w = 1800$ and $L_g = 1800$, respectively, while ReIR modeling lengths were $L_a = L_h = 4500$. The desired speech delay was $\Delta = 240$ (\qty{6}{\milli\second}). Internal processing latencies of $2$ and $3$ samples from the dSPACE system were incorporated for the feedforward and feedback paths, respectively. The amplification factor for the desired speech was set to $\alpha = 2.0$. To obtain stable operation in the considered setup, the regularization parameters were chosen relative to the largest eigenvalue ($\lambda_{\max}$) of the input correlation matrix. In particular, the feedforward regularization parameter was set to $\beta_{\mathrm{FF}} = \lambda_{\max}/10^4$. The feedback channel generally needs more regularization to prevent instability, and thus $\beta_{\mathrm{FB}} = 30\,\beta_{\mathrm{FF}}$ was used. The trade-off parameter $\mu$ was varied from \qty{1e-6}{} to 3000, which corresponds to a $\log_{10}(\mu)$ range of $-6$ to 3.48, to evaluate the performance.

\subsection{Evaluation metrics}
The performance was evaluated in terms of noise reduction, speech distortion, speech quality and intelligibility.

The noise reduction is defined as the difference between the power of the noise component of the leakage $p_v(n)$ (without control) and the noise component of the inner error microphone signal $e_v(n)$ (with control), i.e.,
\begin{equation}
    \mathrm{NR} \ \mathrm{(dB)}  
    = 
        10\log_{10}   \sum\limits_{n=1}^{N}  p_v^2(n) 
        -
        10\log_{10}   \sum\limits_{n=1}^{N} e_v^2(n) ,
\end{equation}
where $N$ denotes the total signal length.

The intelligibility-weighted spectral distortion is used to assess the amount of speech distortion~\cite{Spriet2004spatially, Doclo2007frequency, Serizel2010integrated}. It is defined as
\begin{equation}
    \mathrm{SD}_\mathrm{intellig} \, \mathrm{(dB)}
    = 
        \sum\limits_{{b}=1}^\mathcal{B} I(\omega_b) \, 10\log_{10} \frac{\mathcal{P}_\epsilon (\omega_b)}{\mathcal{P}_{\mathrm{ref},s} (\omega_b)}  ,
    \label{eq:SD_intellig}
\end{equation}
where the band importance function $I(\omega_b)$ expresses the importance of the ${b}$-th one-third-octave band for intelligibility~\cite{ASA1997}, and $\mathcal{B}$ denotes the total number of bands. $\mathcal{P}_\epsilon (\omega_b)$ is the power spectral density of $\epsilon(n)$ in the $b$-th band, where $\epsilon(n) = e_s(n) - \alpha x_{\mathrm{ref},s}(n-\Delta)$. $\mathcal{P}_{\mathrm{ref},s} (\omega_b)$ is the power spectral density of $\alpha x_{\mathrm{ref},s}(n-\Delta)$ in the $b$-th band.

In addition, we considered the narrowband perceptual evaluation of speech quality (PESQ)~\cite{Rix2001PESQ} and the extended short-term objective intelligibility (ESTOI)~\cite{Jensen2016algorithm} metrics using $\alpha x_{\mathrm{ref},s}(n-\Delta)$ as the reference signal. We evaluated the changes in PESQ and ESTOI from the leakage (without control) to the inner error microphone signal (with control), with positive values indicating improvement.

\begin{figure}[t]
    \centering
    \includegraphics[height=0.73\linewidth]{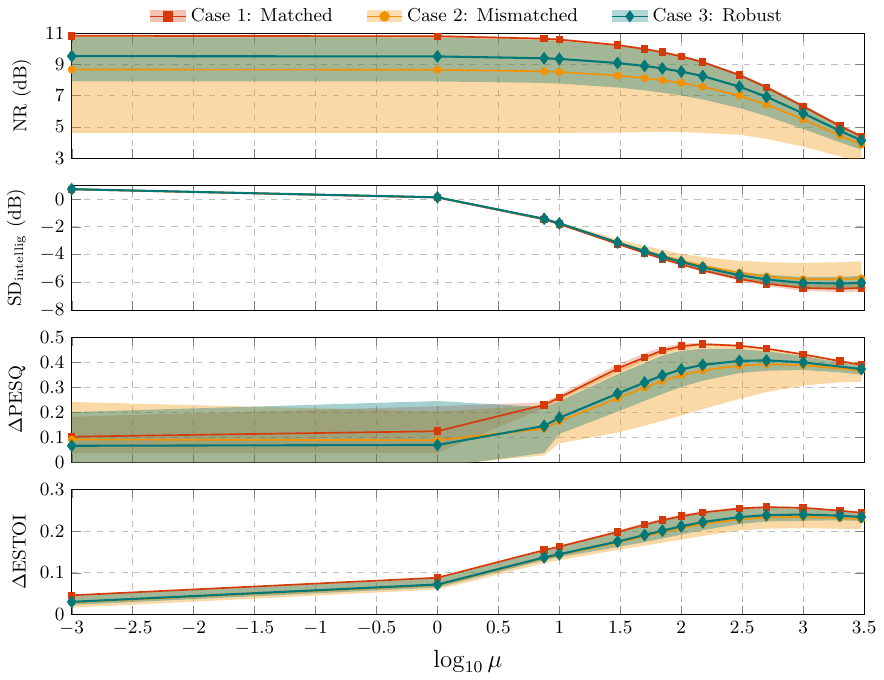}
    \vspace{-12pt}
    \caption{Noise reduction, intelligibility-weighted spectral distortion, narrowband PESQ improvement, and ESTOI improvement for the matched case (Case 1), the mismatched case (Case 2), and the robust case (Case 3) at the right inner error microphone (\#6) with respect to different values of the trade-off parameter $\mu$. Solid lines denote the mean across scenarios, and the shaded regions indicate the 5th--95th percentile range. Results for $\log_{10}(\mu)<-3$ are the same as for $\log_{10}(\mu)=-3$ and are omitted for clarity.}
    \vspace{-6pt}
    \label{fig:result_1}
\end{figure}

\subsection{Simulation results}
The performance of the three cases is illustrated in \cref{fig:result_1} for the right inner error microphone (\#6). The solid lines denote the mean performance across all evaluated scenarios, while the shaded regions indicate the 5th--95th percentile range. The same trends were observed at the left ear (\#5) and are omitted for brevity.

For small $\mu$ values (e.g., $\log_{10}(\mu)<1$), the system essentially behaves as a conventional ANC system, and the PESQ and ESTOI improvements may therefore not be meaningful. For larger $\mu$ values, the matched case generally achieves the highest mean noise reduction, PESQ improvement, and ESTOI improvement, together with the lowest speech distortion, while exhibiting only a narrow spread. In contrast, the mismatched case shows a substantially wider performance range, particularly for noise reduction, where the 5th--95th percentile interval spans up to approximately \qty{6}{\decibel}. The PESQ improvement also exhibits a noticeably larger spread, whereas speech distortion and ESTOI improvement are less affected by the mismatch. The proposed robust case achieves mean performance slightly lower than the matched case but comparable to the mismatched case, while substantially narrowing the 5th--95th percentile range, particularly for noise reduction and PESQ improvement. These results indicate that the robust filter reduces sensitivity to secondary-path variations and provides more consistent performance across the considered range of $\mu$.

\subsection{Experimental validation}
\begin{figure}[t]
    \centering
    \includegraphics[width=0.88\linewidth]{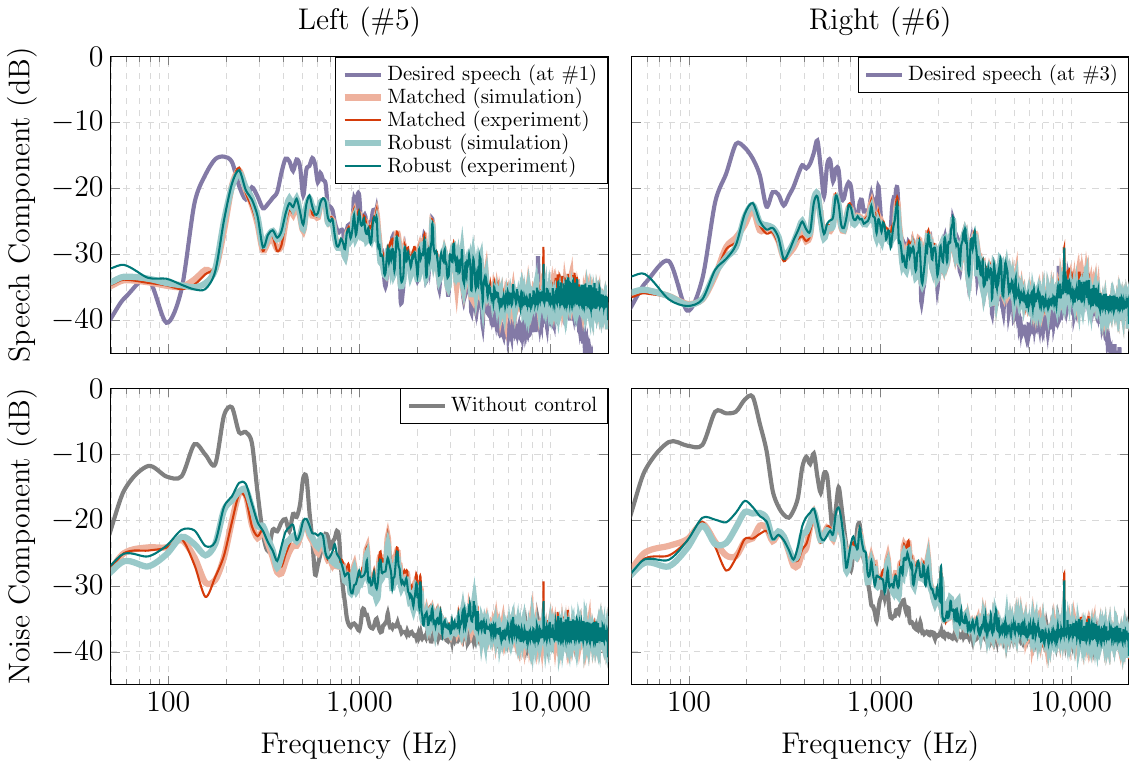}
    \vspace{-14pt}
    \caption{Spectra of the speech component (top row) and the noise component (bottom row) of the inner error microphone signals (\#5 and \#6) from the simulated and experimental evaluations for the matched and the robust cases for $\log_{10}(\mu) = 2.18$.}
    \vspace{-6pt}
    \label{fig:result_2}
\end{figure}

To validate the simulation results, we implemented the control filters obtained from the matched and robust optimizations on the real-time dSPACE system in the same acoustic scenario. The control filter for the matched case was calculated according to \labelcref{eq:w_ssanc_time_soft} using the matched secondary path from KEMAR, and the filter for the robust case was calculated according to \labelcref{eq:w_robust_solution} using the average matrices across all 44 paths. The inner error microphone signals were recorded both without and with control. The spectra of the speech and noise components at the inner error microphones for $\log_{10}(\mu) = 2.18$ are shown in \cref{fig:result_2}. The experimental results are consistent with the simulations for the tested configuration, supporting the validity of the simulation model.

\section{Conclusion}
\label{sec:conclusion}
This paper examined the impact of secondary-path variations on the performance of spatially selective active noise control systems for hearables. Simulation results showed that the matched case, assuming oracle knowledge of the secondary path, generally achieved the best overall performance. However, secondary-path mismatch substantially increased performance variability and degraded performance in some conditions, particularly in terms of noise reduction and speech quality. To address this issue, we proposed a robust optimization framework that minimizes the average cost over a set of secondary-path estimates. The resulting robust filter achieved slightly lower mean performance than the matched case but more consistent performance than filters optimized for a single mismatched path across the considered set of estimates. Experimental results obtained using the tested head-and-torso simulator were consistent with the simulation findings. Future work will investigate the impact of ReIR estimation errors on the performance of spatially selective active noise control systems.

\clearpage

\bibliographystyle{IEEEtran}
\bibliography{refs}

\end{document}